\documentclass[aps,prl,twocolumn,groupedaddress]{revtex4-1}
\usepackage{graphicx}
\usepackage{subfig}
\usepackage[draft]{hyperref}
\usepackage{amsmath}
\usepackage[justification=raggedright]{caption}
\begin{document}
\preprint{APS/123-QED}

\title{Ultra-Low-Intensity Magneto-Optical and\\ Mechanical Effects in Metal Nanocolloids}

\author{M. Moocarme$^{1,2}$}
\email{mmoocarme@qc.cuny.edu}
\author{J. L. Dominguez-Juarez$^2$}
\author{L. T. Vuong$^{1,2}$}
\email{Luat.Vuong@qc.cuny.edu}
\affiliation{$^1$Physics Department, The Graduate Center of CUNY, 365 5th Ave, New York, NY, USA}
\affiliation{$^2$Physics Department, CUNY Queens College, 65-30 Kissena Blvd, Flushing, Queens, NY, USA}
\date{\today}

\begin{abstract}
We theoretically and numerically investigate the linear and nonlinear magneto-optical contributions to the refractive index of metal nanospheres. The analysis is in good agreement with the experimental extinction spectra of gold nanocolloid solutions, with threshold magnetic fields less than 1 mT when illuminated with light intensities less than 1 W/cm$^2$. Plasmonic current loops and vortex power flows provide a new framework for the dynamical interaction between material magnetization, light polarization and nano-surfaces. The photo-induced magneto-optical interaction of non-magnetic metal nanocolloids impart significant, non-negligible torque forces in nanofluids.
\end{abstract}


\maketitle

\begin{figure}[t]
\centering
\includegraphics[width=0.48\textwidth]{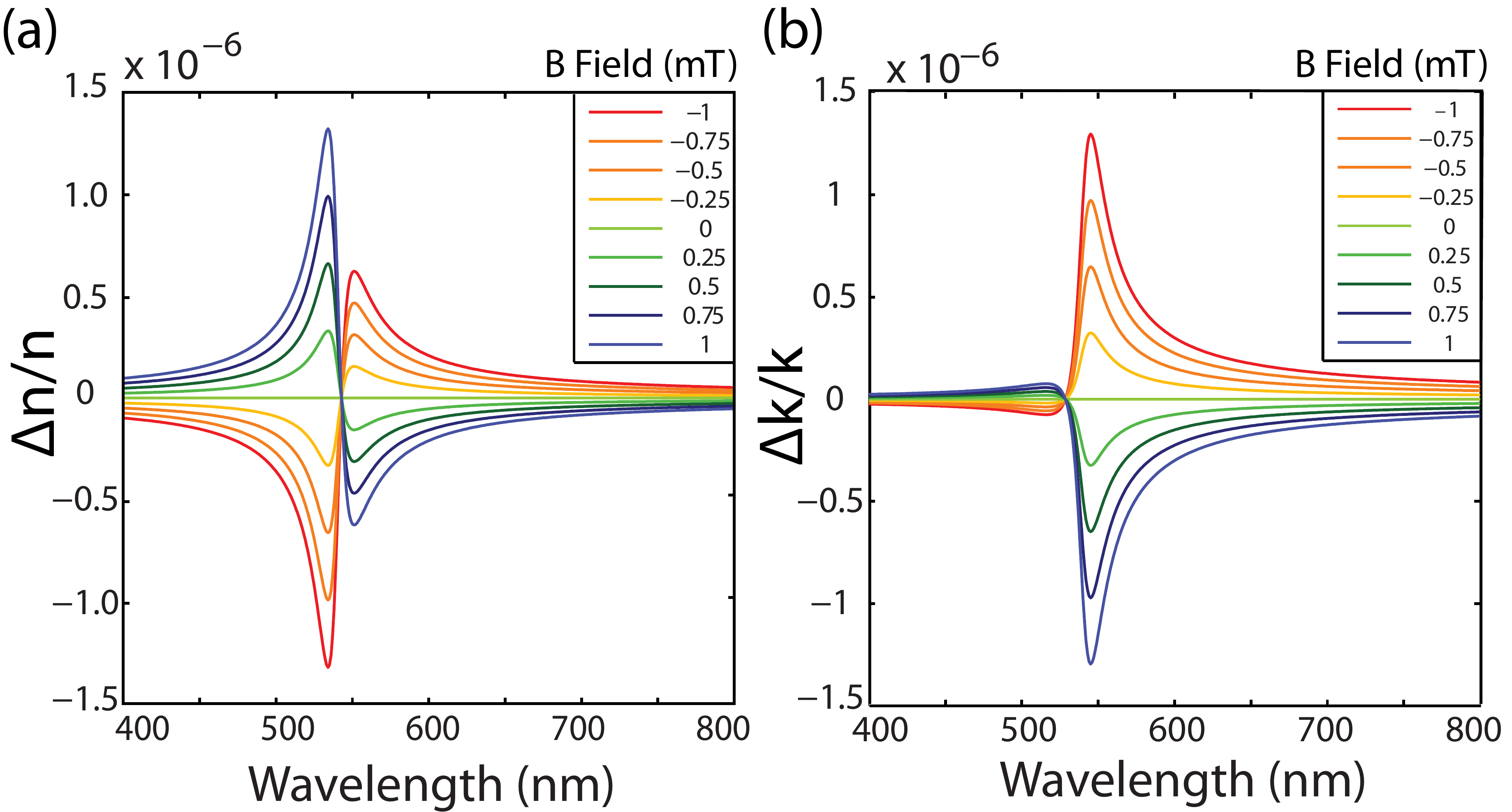}
\caption{Relative MO shift of bulk gold when varying external fields. (a) Real and (b) imaginary parts of the refractive index as a function of wavelength under illumination of 1 W/cm$^2$ RHCP with varying magnetic fields.}
\label{fig:1}
\end{figure}

\begin{figure}[b!]
\centering
\includegraphics[width=0.48\textwidth]{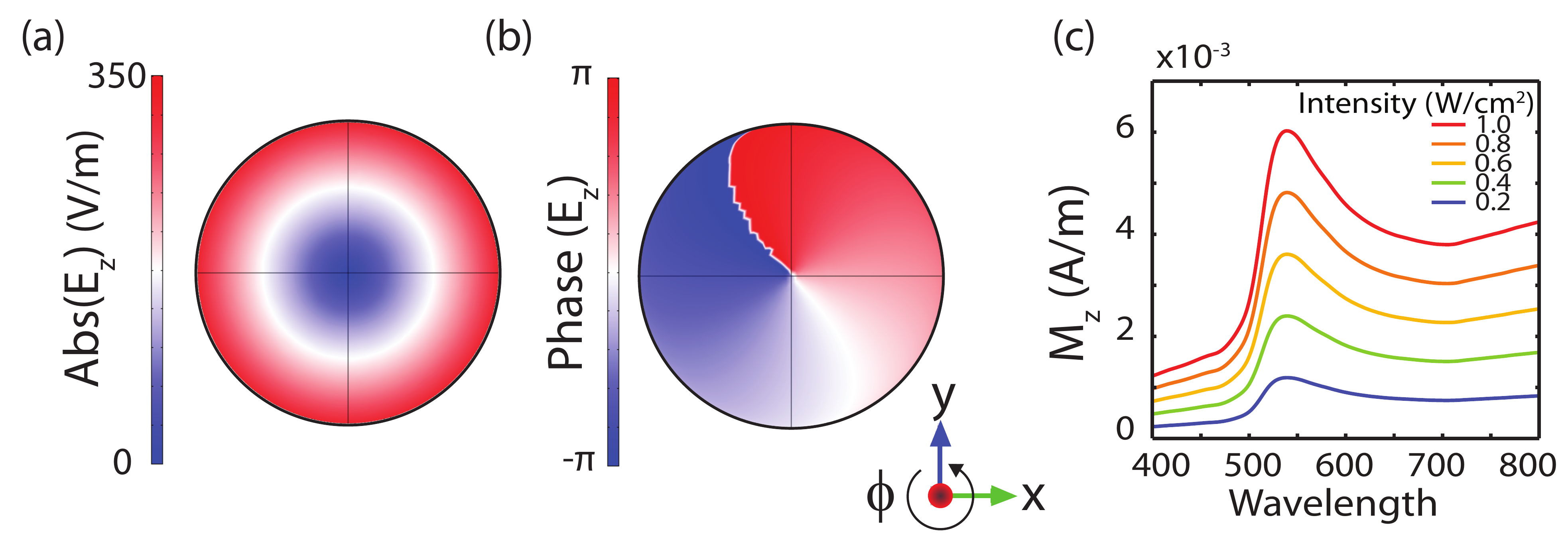}
\caption{Numerically calculated (a) amplitude and (b) phase of the longitudinal component of the electric field, $E_z$, with the application of 1W/cm$^2$ illumination intensity. (c) Volume-averaged magnetization as a function of wavelength at varying intensities. All plots use an aqueous 80-nm gold nanoparticle with application of RHCP.}
\label{Ez}
\end{figure}

\begin{figure}[t]
\centering
\includegraphics[width=0.48\textwidth]{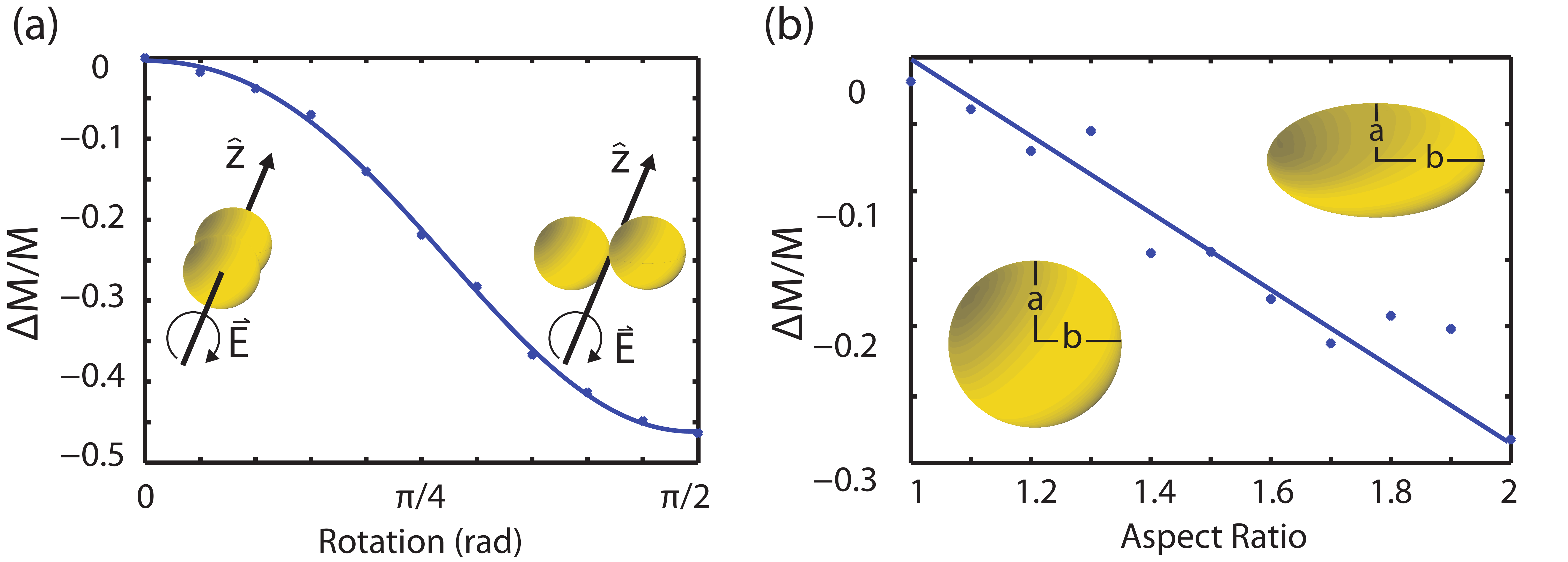}
\caption{Numerically calculated relative change in magnetization (a) of a 80-nm gold dimer nanocluster as a function of angle of rotation, and (b) rotating an ellipsoidal gold nanoparticle from minimal to maximal surface area as a function of aspect ratio, illuminated with 1 W/cm$^2$ RHCP.}
\label{Dimer}
\end{figure}

The plasmonic resonances of noble metal nanostructures manifest in the visible spectral range and have a fundamental role in shaping the optical properties of materials~\cite{Link,Cortie}. The high concentration of electromagnetic fields at a sub-wavelength structure~\cite{Barnes,Kelly,Ozbay} is crucial to numerous methods that modify a materials refractive index, i.e. electro-optically~\cite{Dicken,Chyou} or thermally~\cite{Nikolajsen}. Moreover, if the material exhibits magneto-optical (MO) behavior, the concurrent application of a magnetic field produces additional optical responses such as the enhancement of the Faraday rotation~\cite{Du}, the introduction of new magnetic modes~\cite{Tang,Alu,Fan}, and the MO enhancement of localized and propagating surface plasmons~\cite{Sepulveda,Yang}. In fact, the polarization-dependent excitation of surface plasmon polaritons (SPPs) via magnetic induction currents is not only highly efficient, but the resulting plasmons are potentially comparable in strength to those excited by electric fields~\cite{Lee}. Furthermore, the plasmon-assisted modulation of the refractive index is achieved with the application of magnetic fields~\cite{Temnov,Gonzalez,Deng}. Correspondingly, ferromagnetic materials are often mated with noble metals to achieve the MO response. In non-magnetic media, the inverse Faraday effect is a polarization-dependent MO effect that leads to the formation of DC magnetic dipoles~\cite{Hertel}. The response is claimed to be too small to observe at room temperature~\cite{Gu} (the effect is observed with high intensity lasers, $>$370 MW/cm$^2$~\cite{Raja}).
\\\indent Here, we identify MO responses that occur when elliptically-polarized light and DC magnetic fields are coincident on non-magnetic metal nanoparticles. The response is observable using ultra-low illumination intensities ($\leq 1$ W/cm$^2$) and mT DC magnetic fields. Moreover, we show that there exists a nonlinear response that is explained with the theoretical model of Ref.~\cite{Hertel} when applied to nanostructures. DC magnetic dipoles arise from the coupling between scattered and incident fields~\cite{Singh, Brandao, Bliokh} where vortex phases characterized by phase singularities are the signature of azimuthal power flow~\cite{Vuong, Hasman}. The vortex flows studied here differ distinctly in character from polar-coordinate whirlpool flows investigated in other work~\cite{Boriskina,Bashevoy}. We claim that the DC magnetic dipoles that are non-negligible and capable of influencing nanocolloids optical properties using external magnetic fields. %, despite reasons that the response should be difficult to observe in non-magnetic media~\cite{Gu}. 
In prior work, we investigated the scattering and describe the low-intensity, nonlinear, broadband MO responses in aqueous gold nanospheres that occur on minute timescales~\cite{Singh}. 
\\\indent In this Letter, we demonstrate that both linear and nonlinear plasmon dynamics lead to MO responses. We distinguish characteristics associated with volume charge densities in the linear regime and surface charges associated with SPPs in the nonlinear regime that dominate the response above low-threshold magnetic fields. In the former, a conventional Hall-effect Drude model~\cite{Mulvaney} provides magnetic field dependent refractive index. In the latter the inverse Faraday effect yields a nonlinear current density proportional to the incident intensity of electric field~\cite{Hertel}. It is fundamental to our analysis that while there is no net current i.e., $\langle\vec{j_x}\rangle= \langle\vec{j_y}\rangle= \langle\vec{j_z}\rangle=0$, non-zero time-averaged current loops exist such that $\langle\vec{j_{\phi}}\rangle \neq 0$. This existence of $\langle\vec{j_{\phi}}\rangle$ forms the basis of the MO response. Our theoretical results are demonstrated via numerical simulations, as well as via experiments. We observe increased absorption and scattering \cite{Singh} of nanocolloid solutions generated from the coincident circularly-polarized light and application of DC magnetic fields. 
\\\indent In the remainder of this Letter we provide analytical and numerical calculations that predict the formation of DC magnetic dipoles when elliptically-polarized light illuminates nanoparticles. We attribute the minute-time responses to the movement and alignment of the photo-induced magnetic dipoles generated on the nanoparticles with the external magnetic field. %Simulations confirm that asymmetry and irregularities in the shapes of the nanoparticles, as well as dimer nanocluster aggregations, produce differences in the induced magnetization depending on the orientation with incident light.%
\\\indent In the linear regime the volume charge density, within the quasi-static limit~\cite{Maier}, is governed by the equations of motion: $m^*\frac{d^2\vec{r}}{dt^2}+m^*\gamma\frac{d\vec{r}}{dt} = q\vec{E} + q(\frac{d\vec{r}}{dt}\times\vec{B})_r-k\vec{r}$, where $\vec{r}$ denotes the spatial coordinate, $m^*$ is the effective mass of the electron, $\gamma$ is the decay rate, $q$ is the charge of an electron, and $k$ is the force constant associated with the restoring force. The electric field propagates in the $\hat{z}$ direction with time-harmonic circular-polarization of angular frequency $\omega$ and amplitude $E_0$, $\vec{E_{\pm}}=E_0 e^{-i\omega t}(\hat{x}\pm i\hat{y})$, where $+(-)i\hat{y}$ represents a right(left)-handed circular-polarized wave or RHCP(LHCP). When an externally-applied DC magnetic field, is aligned with the direction of the electric-field propagation, $\vec{B}=B_0\hat{z}$, the system of equations yield the dielectric function as a gyrotropic tensor $\epsilon=1+ \omega_p^2\Bigl(\begin{smallmatrix} \alpha& -\beta &0\\ \beta &\alpha&0\\0&0&\zeta\end{smallmatrix}\Bigr)$, in which $\alpha=\frac{\omega_0^2-\omega^2-i\gamma\omega}{(\omega_0^2-\omega^2-i\gamma\omega)^2+(\omega\omega_c)^2}$, $\beta=\frac{i\omega\omega_c}{(\omega_0^2-\omega^2-i\gamma\omega)^2+(\omega\omega_c)^2}$, and $\zeta=\frac{1}{\omega_0^2-\omega^2-i\gamma\omega}$, where $\omega_c=qB_0/m$ is the cyclotron frequency, $\omega_p$ is the plasma frequency such that $\omega_p^2 = \eta q^2/m\epsilon_0$, $\omega_0$ is the frequency associated with a harmonic restorative force, and $\eta$ is the electron density. 
Figure~\ref{fig:1} shows the shifts in the real and imaginary parts of the refractive index with significant modulation around the plasmon resonance at 540 nm. 
 The MO model shown above [Fig.~\ref{fig:1}] describes the response due to magnetic field perturbations as well as our experimental samples while "settling", however is insufficient in explaining changes in the scattering spectra at higher applied magnetic fields~\cite{Singh}. %One possible reason for deviation at higher magnetic fields is that our theory considers individual nanoparticles in a surrounding medium, whereas ensemble averages of interactions between the metal spheres make significant differences in the nonlinear surface effects~\cite{Garcia-V}. 
\begin{figure}[b!]
\centering
\includegraphics[width=0.5\textwidth]{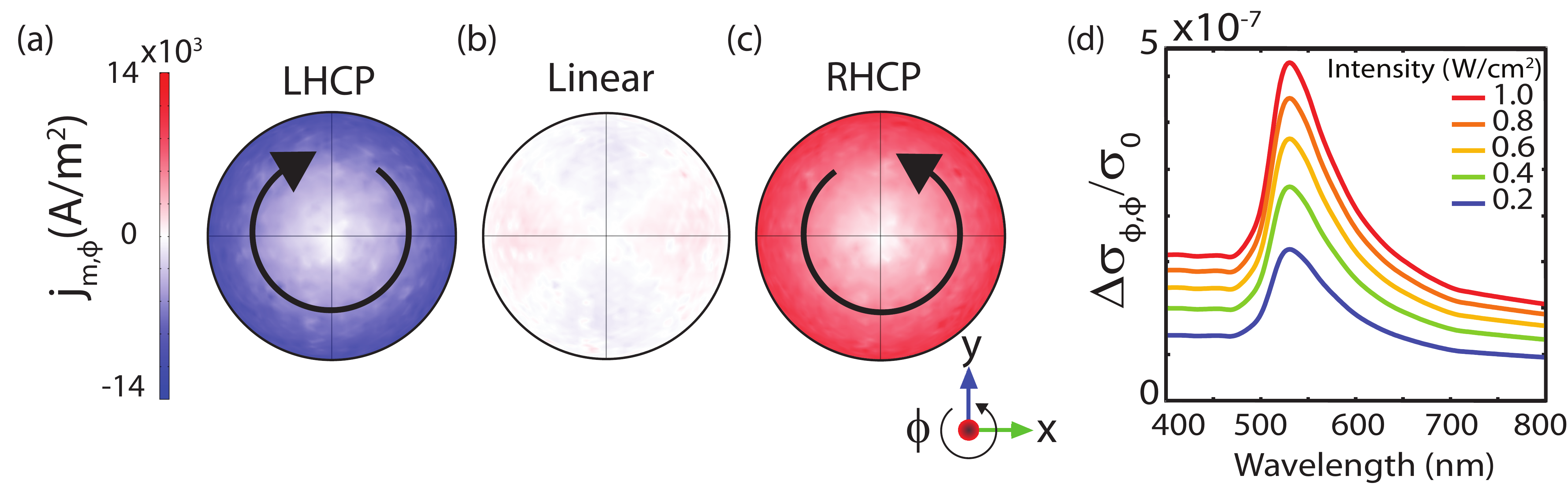}
\caption{(a) Surface plots of the nonlinear current density of an aqueous 80-nm gold nanosphere when illuminated by LHCP, linearly-polarized and RHCP light in the $\hat{z}$ direction at the plasmon resonance, 540 nm, at 1W/cm$^2$. (b) Volume-averaged relative change in azimuthal component of the conductivity as a function of wavelength with varying illumination intensity for RHCP or LHCP.}
\label{fig:2}
\end{figure}
\\\indent In departure from prior investigations of MO effects on nanostructures, we use the analytical expression from~\cite{Hertel} to describe a time-averaged solenoidal magnetization current density derived from the continuity equation:
\begin{equation}
\langle \vec{j_m} \rangle = \frac{i}{4q\langle \eta\rangle \omega}\nabla \times\big(\sigma_0^*\vec{E}^*\times\sigma_0\vec{E}\big).
\label{jm_eqn}
\end{equation}
Where $\langle \eta\rangle$ is the time-averaged electron density and $\sigma_0$ is the bulk conductivity of gold. For the case of a circularly-polarized wave propagating in the $\hat{z}$ directions the term $\vec{E_{\pm}}\times\vec{E_{\pm}^*}$ can be written as $\pm i |\vec{E_0}|^2\hat{z}$. Note that $\langle\vec{j_m}\rangle$ scales inversely with $\langle \eta\rangle$, which may explain MO effects in non-metallic nanostructures~\cite{Kuznetsov}. $\langle\vec{j_m}\rangle$ also scales with the intensity, whereas the linear currents, derived from a conventional Hall-effect model, scale with the electric field ($\vec{j}=\sigma_0\vec{E}$). $\langle\vec{j_m}\rangle$ exhibits polarization dependence in direction and magnitude; for incident linearly-polarized light $\langle\vec{j_m}\rangle$ is strictly $0$, is non-zero for any degree of elliptical polarization, and greatest at either RHCP or LHCP. In our investigation of nanospheres, the current density is most significant in the azimuthal direction, $\hat{\phi}$, $\langle\vec{j_{m,r}}\rangle=0$, $\langle\vec{j_{m,\theta}}\rangle\ll \langle\vec{j_{m,\phi}}\rangle$, and subsequently, the induced azimuthal surface currents are an anisotropic perturbation to the conductivity of the nanoparticle. The approximation $\vec{j_i}=\sigma_{ij} \vec{E_j}$ yields an expression for the relative change in the azimuthal component of the conductivity tensor:
\begin{equation}
\frac{\Delta\sigma_{\phi,\phi}}{\sigma_0} = \frac{i|\sigma_0|}{4q\langle\eta\rangle\omega}\frac{\big(\nabla\times(\vec{E}^*\times\vec{E})\big)_{\phi}}{\vec{E}_{\phi}}.
\label{DelSig}
\end{equation}
\\\indent We complement the analytical expression [Eq.~\ref{DelSig}] with numerical simulations, using finite-element method software. Figure~\ref{fig:2}(a) show the calculated nonlinear current density on an aqueous 80-nm gold nanosphere illuminated in the $\hat{z}$ direction with different polarizations. The numerical calculations illustrate that RHCP and LHCP yield the greatest $\langle\vec{j_m}\rangle$ with equal magnitude and opposite direction, while $\langle\vec{j_m}\rangle$ is negligible for incident linear polarization. The relative change in azimuthal conductivity [Eq.~\ref{DelSig}] is illustrated in Fig.~\ref{fig:2}(b). Increasing the light intensity results in the increase in the conductivity of the nanoparticle uniformly across wavelengths for both RHCP and LHCP. Both the azimuthal currents and the electric-field components reverse direction with polarization handedness; by symmetry, the conductivity is identical for both RHCP and LHCP. 
\begin{figure}[t]
\centering
\includegraphics[width=0.48\textwidth]{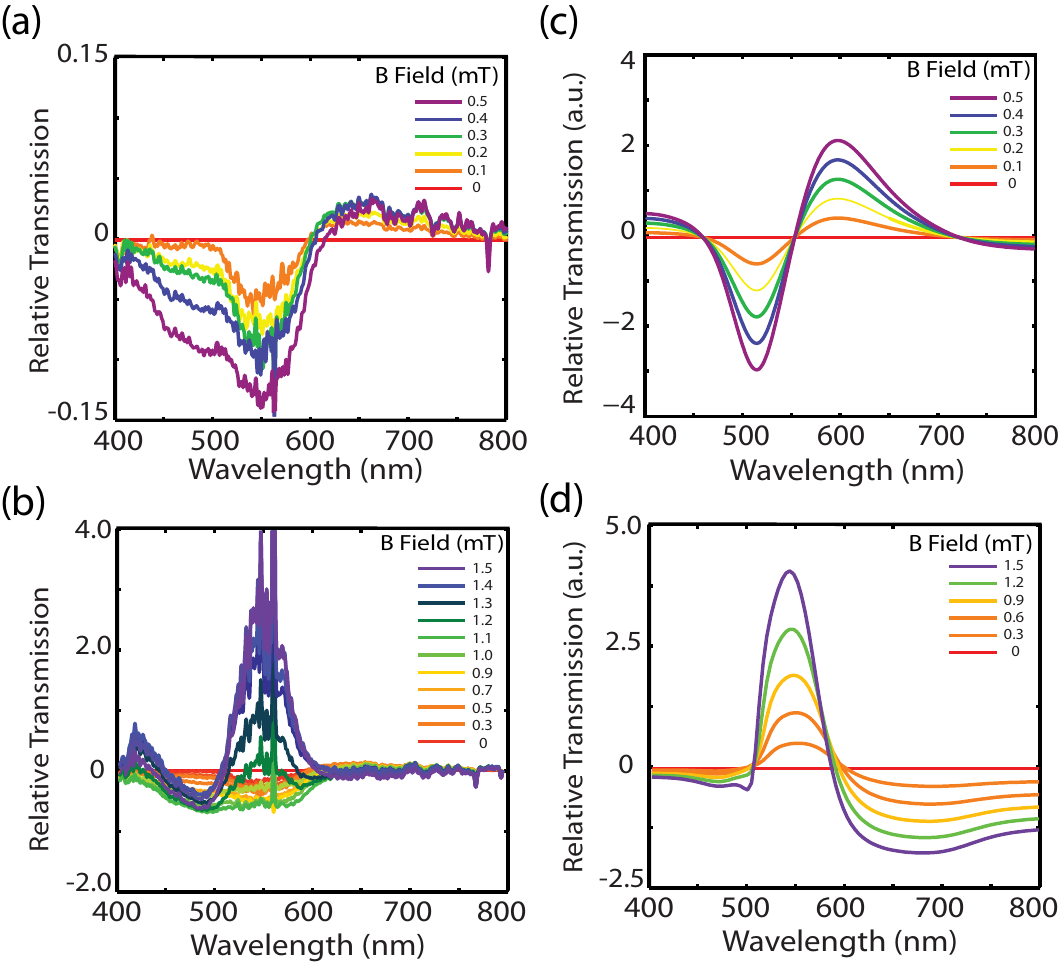}
\caption{Normalized MO changes in transmission for aqueous gold nanocolloid at (a) low magnetic fields and (b) higher magnetic fields. Theoretical MO change in transmission of aqueous 80-nm gold nanospheres in the (c) linear and (d) nonlinear analyses. In all plot the illumination is 1 W/cm$^2$ and RHCP.}
\label{Exp}
\end{figure}
\\\indent Experimental data is illustrated in Fig.~\ref{Exp}, alongside analytical results. Samples of 0.05 mg/mL 80-nm poly-vinyl pyrolidone(PVP)-capped gold nanoparticles are dispersed in aqueous solution (2.5$\times 10^{-6}$ fill factor). A solar simulator, polymer thin-film linear polarizer, and 400-800 nm achromatic quarter-wave plate produce RHCP white light, and a Helmholtz coil generates a uniform magnetic field in the direction of propagating light around the sample. A CCD spectrometer that captures the extinction spectra is placed behind a 0.5 cm optical path length cuvette containing samples.
\\\indent The experimental data [Fig.~\ref{Exp}] shows relative transmission data for RHCP and positive magnetic fields, however the transmission is similar for both RHCP and LHCP above threshold magnetic fields of 1 mT. Below the threshold magnetic field differences between the handedness of the polarizations emerge, the response is equal and opposite in magnitude. 
\\\indent Figures~\ref{Exp}(c,d) are calculated by incorporating the changes of the refractive index [Fig.~\ref{fig:1}] and conductivity [Fig.~\ref{fig:2}(e)] to Mie theory~\cite{Mie} and computing the relative changes in extinction spectra. We model the nonlinear response as exhibiting a dependence on the external magnetic field by assuming that larger MO responses result from greater magnetic fields. Experimentally, there is a clear shift in the extinction spectra above and below 0.9 mT, which we associate as the bound between the linear and nonlinear regimes. 
\\\indent Both the linear and nonlinear MO effects studied coincide with polarization-dependent light scattering from metallic nanostructures; the scattered or longitudinal component of the electric field $E_z$ scales with the induced charge density $\tilde{\rho}$ in the limit of thin structures. Figure~\ref{Ez}(a) shows the amplitude of the scattered field, which couples to the incident field to produce nonlinear currents [Eq.~\ref{jm_eqn}]. The nonlinear azimuthal surface currents can be interpreted as the response of $|\tilde{\rho}|$ from incident azimuthal electric fields. Figure~\ref{Ez}(b) shows the phase singularity of $E_z$ at the surface of an aqueous 80-nm gold nanoparticle scattered by incident RHCP for $\lambda = 540$ nm. The vortex phase of $E_z$ changes direction with incident polarization handedness \cite{Vuong} in a manner similar to the azimuthal current density in the linear regime. Figure~\ref{Ez}(c) shows the volume-averaged magnetization induced on an 80-nm aqueous gold nanoparticle, which follows a similar trend to the current density, exhibiting a peak response at the plasmon resonance. Although the nonmagnetic nanoparticle magnetization is small compared to that of a ferromagnet($10^{5}$ times smaller), due to the small mass, mechanical forces such as dipole-dipole interactions and torques from an external magnetic field can be significant.%the resulting torque produces significant angular accelerations ($\approx$10$^6$ rad/s$^2$).
\\\indent Changes in the extinction spectra are associated with MO shifts in the refractive index, however the minute-time response to the applied magnetic fields is identified with the motion of the nanoparticles that may be attributed to the alignment of nanoparticles and nanocluster aggregates. Similarly, we acknowledge that the nanoparticles employed in experiments are not perfect spheres, but exhibit irregularities from the fabrication process; such variations from spherical geometry cause magnetization differences that depend on the nanoparticle orientation with respect to the electric field. In this perspective, it is not that the magnetization of the nanoparticles is necessarily observed directly, but the magnetization associated with nanoparticle and nanocluster reorientation.
\\\indent Numerical calculations of dimer nanoclusters and ellipsoids of different aspect ratios identify how the structural anisotropy affects the magnetization of the nanoparticle. Figure~\ref{Dimer}(a) illustrates that dimer nanoclusters show a reduction in the magnetization by almost 50\% as the nanocluster rotates from minimal to maximal incident surface area with respect to electric field.
 Figure~\ref{Dimer}(b) shows the relative change in the magnetization as a function of aspect ratio when rotating ellipsoidal nanoparticles of equal volume from minimal to maximal incident surface area. Ellipsoids with aspect ratio 1 (spheres), exhibit no difference as they are rotated, and greater differences are observed with increasing aspect ratio. To guide the eye, a trendline is provided in Fig.~\ref{Dimer}(b), reflecting the Biot-Savart law. Varying the shape of the nanoparticle or nanocluster causes a shift in the plasmonic eigenfrequency, which influences the trend associated with the nanoparticle magnetization. Regardless, anisotropy in the nanoparticle shape leads to MO magnetization that varies with the orientation, a concept seen in other plasmonic responses of nanoparticles~\cite{Jain,Funston}. 
\\\indent %Numerical simulations of dimer nanoclusters and ellipsoids indicate that structural anisotropy lead experimentally to torque forces that maximize MO and mechanical effects.
There is an optimal orientation for the nanoparticles to align with the electric field, because differences in structure result in variations in the induced magnetization that are orientation-dependent.
% Anisotropy on the same order as ellipsoids simulated give rise to minute-time scale responses observed. \\
%\begin{figure}[t]
%\centering
%\includegraphics[width=0.48\textwidth]{ChangeNLMagAspectRatioandRotation2}
%\caption{Relative change in magnetization from rotating from minimal to maximal surface area of an ellipsoidal nanoparticle as a function of aspect ratio.}
%\label{Ellip}
%\end{figure}
%Simulations indicate that the optimal orientation for the nanoparticles to align with the electric field is for minimal surface area incidence.
The application of an external magnetic field increases nanocolloid MO responses by changing the alignment of nanoparticles. By exciting a magnetic dipole with the electric field and applying an external magnetic field the system stabilizes in an orientation that maximizes the nanoparticle magnetization, which occurs when the local electric field is largest.
%Nanoparticle magnetization is greatest when the electric and magnetic fields are co-aligned and the orientation of the generated magnetic dipole is reversed upon changing circular-polarization handedness. 

%Simulations indicate that by changing the polarization handedness the direction of the magnetic dipole is reversed, thus, reversing the magnetic field yields similar changes in the nonlinear regime. 
%\\\indent Experimental and analytical data suggest that at high magnetic fields the transmission behaves similarly for both RHCP and LHCP, differences occur at lower magnetic fields where the nonlinear effect isn't as prominent. 
%Particular orientations of aspherical nanoparticles and nanoclusters can accommodate resonances that give rise to the MO response, though this has not been probed in detail in this Letter. %	Pure gold exists as a weakly diamagnetic material, creating a magnetic field opposing the external field. The current loops excited also act to oppose the external field according to Lenz's law, therefore the effect of any diamagnetic properties only increase the MO response.
\indent In general, the MO changes in extinction spectra are larger than anticipated for a given fill factor and we attribute the anomalously-large response to multiple scattering events. The scattering cross-sections of metal nanocolloids are much greater than their geometrical cross-section at resonance~\cite{Mie}. Autocorrelation data shows significant pulse broadening through 400$\mu m$ of nanocolloid solution from 2.3ps to 3.2ps, which suggest appreciable multiple scattering events, despite low fill factors, (2.5$\times 10^{-6}$). In fact, MO responses are observed with linearly-polarized and even unpolarized light \cite{Singh}, though the effects are theoretically predicted to be much smaller in magnitude and sensitive to changes in nanoparticle shape and aggregation. %Multiple-scattering models have shown that reflection and transmission of light from spherical particles are considerable despite low fill factors~\cite{Garcia-V}, and that near-fields around plasmonic nanoparticles dominate the total field within a few wavelengths~\cite{Sondergaard}, hence multiple-scattering effects non-negligible in nanocolloids.
\\\indent Our research provides the theoretical framework to further study MO responses in non-magnetic nanostructures and to optically assemble large quantities of disperse, nonspherical nanoparticles using broadband sources. Our work also identifies new methods of magnetically torquing non-magnetic nanostructures, in particular, biological in-situ applications~\cite{Reenen}.
\\\indent %In conclusion, we develop a theoretical model that successfully explains the nonlinear MO plasmon dynamics of metal nanoparticles when excited by elliptically-polarized electric fields and DC magnetic fields. %Linearly-polarized light may also torque non-spherical particles in this model. 
%Our research indicates that when coincident, such fields magnetize and shift the optical properties of metal nanocolloids. This analysis is in good agreement with experimentally observed extinction and scattering spectra associated with MO and novel mechanical effects in non-magnetic media that were previously considered unattainable. 
In conclusion, metal nanocolloids magnetize and the optical properties are shifted when excited by coincident elliptically-polarized electric fields and DC magnetic fields. In good agreement with experimentally observed extinction and scattering spectra MO and novel mechanical effects are observed in nonmagnetic nanocolloids that were previously considered unattainable. We demonstrate and provide a theoretical model that successfully explains the nonlinear MO plasmon dynamics of metal nanoparticles.
\bibliography{NLNPbib}
\end{document}